\documentclass[a4paper]{article}
\usepackage{a4wide,multicol,xcolor,graphicx,url,orcidlink,thumbpdf,lmodern}
\usepackage[authoryear,round]{natbib}
\usepackage[helvet]{sfmath}

\hypersetup{%
hyperindex,%
colorlinks,%
linktocpage,%
plainpages=false,%
linkcolor=Blue,%
citecolor=Blue,%
urlcolor=Red,%
pdfstartview=Fit,%
pdfview={XYZ null null null}%
}

\definecolor{Red}{rgb}{0.5,0,0}
\definecolor{Blue}{rgb}{0,0,0.5}
\definecolor{hellgrau}{rgb}{0.55,0.55,0.55}

\begin{document}

{\large\bfseries Letter to the editor of \emph{Respiratory Medicine}:}

\bigskip

{\LARGE \bfseries Refuting ``Debunking the GAMLSS Myth: Simplicity Reigns in Pulmonary Function Diagnostics''\par}

\bigskip

\begin{multicols}{3}

\textbf{Robert A. Rigby}~\orcidlink{0000-0003-3853-1707}\\
\emph{University of Greenwich}

\textbf{Mikis D. Stasinopoulos}~\orcidlink{0000-0003-2407-5704}\\
\emph{University of Greenwich}

\textbf{Achim Zeileis}~\orcidlink{0000-0003-0918-3766}\\
\emph{Universit\"at Innsbruck}

\end{multicols}
\begin{multicols}{3}

\textbf{Sanja Stanojevic}~\orcidlink{0000-0001-7931-8051}\\
\emph{Dalhousie University}\\

\textbf{Gillian Heller}~\orcidlink{0000-0003-1270-1499}\\
\emph{University of Sydney}

\textbf{Fernanda de Bastiani}~\orcidlink{0000-0001-8532-639X}\\
\emph{Federal University of Pernambuco (UFPE)}

\end{multicols}
\begin{multicols}{3}

\textbf{Thomas Kneib}~\orcidlink{0000-0003-3390-0972}\\
\emph{Georg-August-Universit\"at G\"ottingen}

\textbf{Andreas Mayr}~\orcidlink{0000-0001-7106-9732}\\
\emph{Philipps-Universit\"at Marburg}

\textbf{Reto Stauffer}~\orcidlink{0000-0002-3798-5507}\\
\emph{Universit\"at Innsbruck}

\end{multicols}

\textbf{Nikolaus Umlauf}~\orcidlink{0000-0003-2160-9803}\\
\emph{Universit\"at Innsbruck}

\bigskip
\bigskip

\emph{Dear Editor,}

We read with interest the above article by \cite{Zavorsky2025} concerning
reference equations for pulmonary function testing. The author compares
a Generalized Additive Model for Location, Scale, and Shape (GAMLSS),
which is the standard adopted by the Global Lung Function Initiative
(GLI), with a segmented linear regression (SLR) model, for pulmonary
function variables. The author presents an interesting comparison;
however there are some fundamental issues with the approach. We welcome
this opportunity for discussion of the issues that it raises.

The author's contention is that (1) SLR provides ``prediction accuracies
on par with GAMLSS''; and (2) the GAMLSS model equations are
``complicated and require supplementary spline tables'', whereas the SLR
is ``more straightforward, parsimonious, and accessible to a broader
audience''. We respectfully disagree with both of these points.

Pulmonary function reference equations are designed to identify
individuals who may have measured values outside what is usually
expected in healthy individuals, i.e., at or beyond the outer centile
curves. This is where \emph{distributional regression} models (e.g.,
GAMLSS) have an advantage compared with the classical regression models
(e.g., SLR). In distributional regression models, multiple features of
the distribution of the response (e.g., \emph{mean}, \emph{variance},
\emph{skewness}, \emph{tail heaviness}) can be modeled as functions of
the explanatory variable(s); and a variety of response distributions of
varying shapes may be assumed. In contrast, in classical regression
models, only shifts in the \emph{mean} of the response variable are
modeled; and the underlying assumption is that the response variables
are normally distributed, with constant variance (i.e.~homoscedastic).

This letter will show that, while the SLR analysis adequately fits the
mean of the response variable, it is not adequate for fitting the tails
of the distribution where, for pulmonary function measures, the focus
should be. The primary reason for this is that SLR models a skew
response variable with a symmetric distribution, leading to poor tail
centiles. We demonstrate that the author's GAMLSS models (Zavorsky
supplementary material Tables S1 and S2) are superior to his SLR models
at identifying respiratory problems.

The identification of people with pulmonary variable values outside the
range usually expected in health is done in two stages.

\begin{itemize}
\item
  \textbf{Stage 1}: Appropriate models are fitted to three responses
  variables, FEV$_1$/FVC, FEV$_1$ and FVC, against the explanatory
  variables height, weight and age, taking possible interactions into
  account.
\item
  \textbf{Stage 2}: The z-scores from the fitted models are used to
  identify patients with measured values potentially falling below the
  lower limit of normal (LLN), at either the 2.5\% or 5\% level.
\end{itemize}

Four different methodologies were used by \cite{Zavorsky2025} for comparing
the models:

\textbf{1) Predicted LLN:} Figure 2 of Zavorsky's paper shows the
predicted 5\% LLN for his GAMLSS and SLR models, at the U.S.\ median
height and age. (Note that for the ratio FEV$_1$/FVC, the SLR reduces
to a simple linear model.) The author claims that SLR produces
``prediction and LLN curves akin to those from GAMLSS''. For
FEV$_1$/FVC (top row), the LLN of GAMLSS and SLR, are similar up to
age 50, but then diverge greatly above age 50, for both females and
males. For FEV$_1$ and FVC (middle and bottom rows), the LLN of GAMLSS
and SLR are similar above age 20, but diverge greatly below age 20, for
both females and males.

Figure 1 below assesses the distributional fit of the GAMLSS and SLR
models in the lower tail, relevant for assessment of accurate LLN
detection. The observed percentage of observations below the 5\% LLN
within 7.5 year age intervals is plotted, for both models. These
percentages, which we refer to as \emph{exceedance percentages}, are
crucially informative of the accuracy of LLN detection. To appreciate
whether the model is good at defining the LLN, we would expect 5\% of
observations to fall below the 5th centile. 95\% confidence bands for
the 5\% percentiles are also shown: exceedance percentages lying within
the 95\% bands should be judged as within the bounds of LLN accuracy.
The GAMLSS observed LLN exceedance percentages lie within the 95\%
confidence bands in all cases except one, indicating accurate detection
of LLN exceedance. Some of the SLR observed exceedance percentages,
however, fall well outside these bands: of particular concern are
FEV$_1$/FVC, males over 50 years (severe overdiagnosis);
FEV$_1$/FVC, females over 70 years (overdiagnosis); and FEV$_1$ and
FVC, children (both genders) under 10 years (underdiagnosis).

\begin{figure}[p!]
\centering
\includegraphics[width=\textwidth]{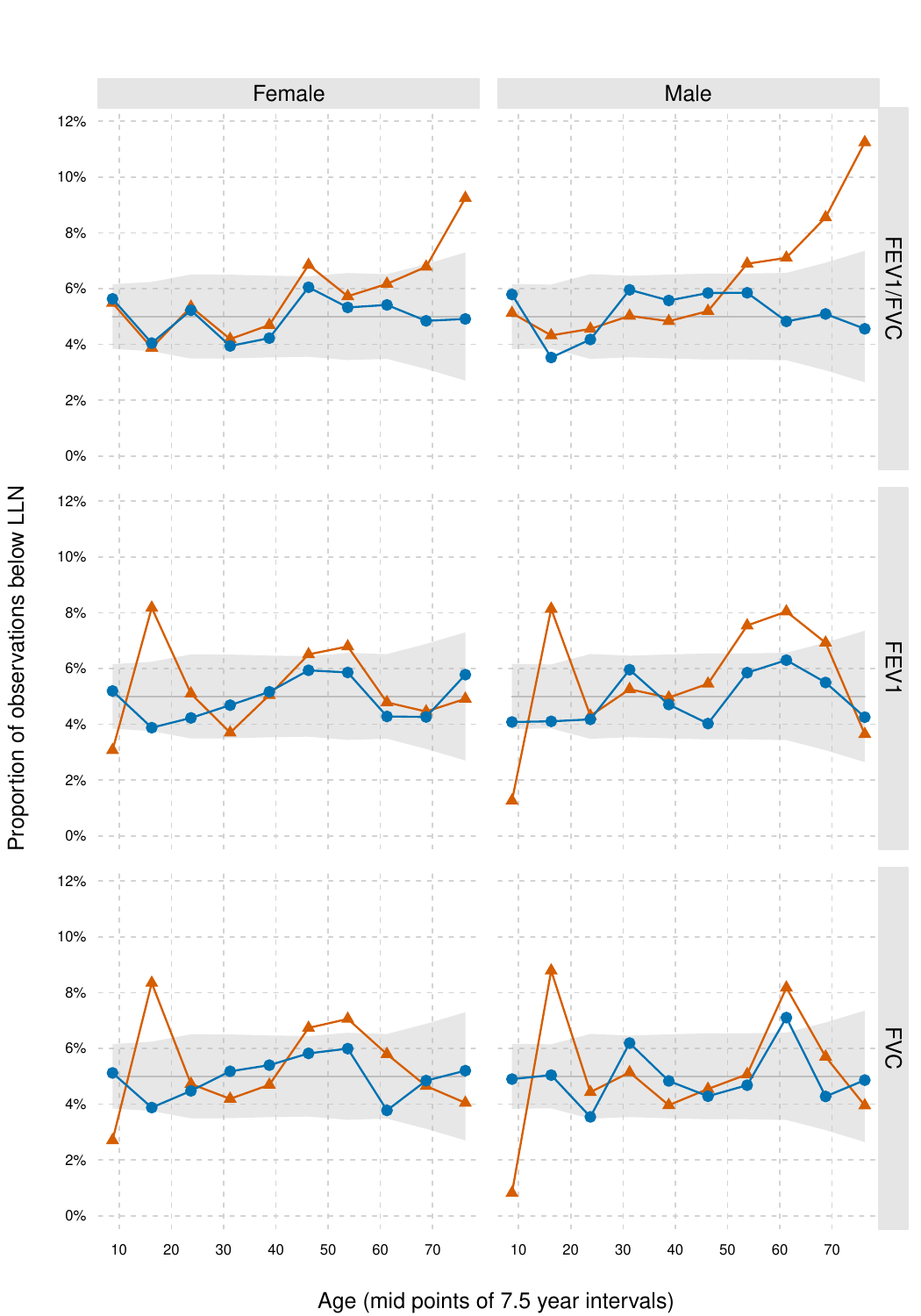}
\caption{Empirical proportions of observations below the 5\% lower limit
of normal (LLN) across different age groups (7.5 year intervals). Red
lines with triangles correspond to the SLR models. Blue lines with
circles correspond to the GAMLSS models. Results are shown for females
(left) and males (right) separately, for each of the variables
FEV$_1$/FVC (top), FEV$_1$ (middle), and FVC (bottom). The
corresponding point-wise 95\% intervals are added in grey in each
panel.}

\end{figure}

\textbf{2) Measures of prediction accuracy:} The predictive performance
in Zavorsky's Table 2 mainly assesses the point accuracy of the
predicted values of the pulmonary variables. It is not informative of
how well the models predict the tail centiles and tail z-scores, where
interest lies. However, the BICs reported in that table already suggest
that the distributional fit of SLR was considerably worse than of
GAMLSS.

\textbf{3) Fitted z-scores:} \cite{Zavorsky2025} compares the z-scores from
the fitted GAMLSS and SLR models (Zavorsky, page 4, bottom of left
column). The use of paired t-tests is inappropriate, because it tests
whether the GAMLSS and SLR z-scores have different population means.
This is not informative about how the z-scores differ in the tails. Also
the fact that all age groups have been combined precludes the discovery
of differences between the GAMLSS and SLR z-scores within different age
groups.

The problematic LLN exceedance percentages for SLR in some age groups,
as identified from Figure 1, correspond to non-normal z-scores. This is
exemplified in Figure 2 below, which shows QQ plots for the SLR and
GAMLSS residuals in one age group with large deviations for each of the
response variables: FEV$_1$/FVC (left), FEV$_1$ (middle), and FVC
(right). Simultaneous confidence intervals (computed with the equal
local levels method) are included and show that the z-scores for GAMLSS
(blue circles) do not deviate too much from normality in all three
subgroups. In contrast, the z-scores for the SLR models (red triangles)
deviate clearly, including large deviations at the 5\% LLNs, highlighted
by gray vertical lines at the theoretical quantiles of $-1.645$. The
QQ plots for females and for other problematic age groups identified
from Figure 1 look qualitatively similar but are omitted here for
brevity.

\begin{figure}[t!]
\centering
\includegraphics[width=\textwidth]{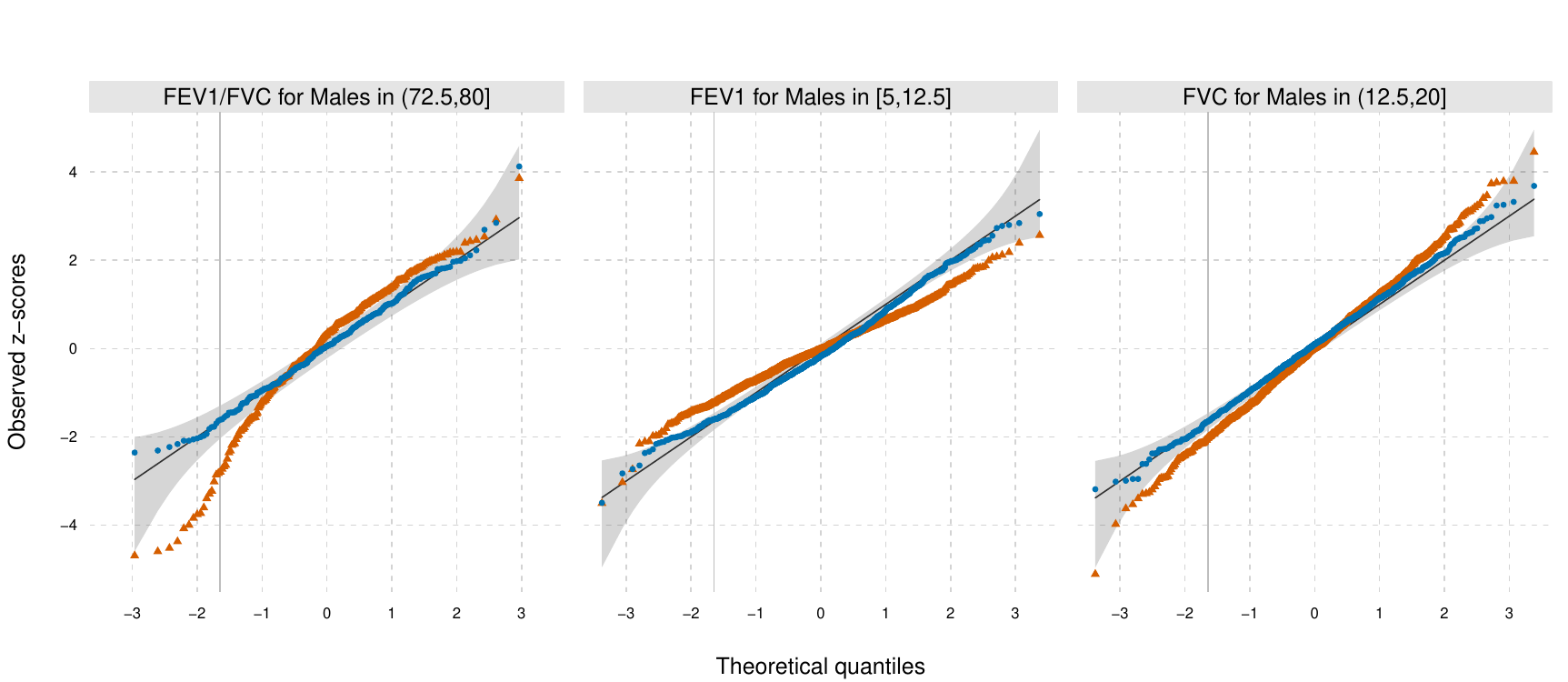}
\caption{QQ plots of the z-scores from the SLR (red triangles) and
GAMLSS (blue circles) models, for a selected age group in each of
FEV$_1$/FVC (left), FEV$_1$ (middle), and FVC (right) response
variables in males.}

\end{figure}

\textbf{4) Classification:} Table 3 of Zavorsky's paper shows that, for
all seven status conditions (for LLN levels 5\% and 2.5\%), there are
significant differences between the population percentages of patients
identified by his SLR and GAMLSS models. The differences in percentages
(and corresponding counts) are often of clinical importance. For
example, in status 3 of the LLN at the 2.5\% level, SLR identifies 3.3\%
(554) of the data cases, while GAMLSS identifies 4.3\% (717) cases. Yet
\cite{Zavorsky2025} concludes (last paragraph of Section 3), that the kappa
statistic indicates ``strong'' or ``moderate'' agreement between the
models. The problem with the kappa statistic is that it provides a
global comparison, not looking at individual or pairs of status
\citep{Maclure1987}. Also this is just a comparison - what matters most is
whether the models accurately discriminate health from disease.

\textbf{Conclusion:} To summarize, the SLR model proposed by \cite{Zavorsky2025}
is inadequate for modeling pulmonary function variables. First,
modeling a skew response variable using a symmetric distribution will
lead to poor tail centile estimates as for example for the FEV$_1$
FVC ratio, where the SLR model uses an inadequate symmetric normal
distribution and the GAMLSS model uses an adequate skew Box-Cox Cole \&
Green distribution. Second, modeling the response variable variance
incorrectly will also lead to poor tail centile estimates, as for
example for FEV$_1$ and FVC, where the SLR model uses an inadequate
two-piece constant variance, while the GAMLSS model uses an adequate
smooth variance function in age. Due to this, the SLR leads to poor
centiles, inaccurate z-scores, and the misclassification of many
patients at risk of respiratory conditions.

As for simplicity, while we acknowledge that GAMLSS model-building is
non-trivial, we find the author's SLR model selection as complicated as
for GAMLSS. \cite{Zavorsky2025} seems to imply that computation of the
GAMLSS models is necessarily manual (``These equations are complicated
and require supplementary spline tables.''). In fact manual computation
is never necessary, and is not recommended as it is error-prone; a GLI
function calculator is available on a public website
(\url{https://gli-calculator.ersnet.org/}) and has been implemented on
most devices and even smartphone apps.

Abandoning the distributional regression approach of GAMLSS, to adopt
the SLR based on the normal distribution, is not advisable when interest
lies in the tails of the response variable distribution. One could look
at ways to simplify the GAMLSS models for the distribution parameters
(e.g., using parametric rather than spline functions, provided they are
adequate), but the author's approach which ignores the distributional
nature of the problem, should not be considered a viable alternative to
the existing GLI standard.

\bibliographystyle{jss}
\bibliography{ref}

\end{document}